\documentclass{aa}
\usepackage{graphicx}
\usepackage{epstopdf}
\usepackage{natbib}
\bibpunct{(}{)}{;}{a}{}{,}
\nonstopmode
\begin{document}

   \title{Transient X-ray Emission from Normal Galactic Nuclei
}

\authorrunning{Wong, Huang \& Cheng}
\titlerunning{Transient X-ray Emission from Normal Galactic Nuclei}

   \author{A. Y. L. Wong\inst{1}, 
                 Y. F. Huang\inst{2}\and
                 K. S. Cheng\inst{1}
                 }

   \offprints{K.S.~Cheng}

   \institute{Department of Physics, Center of Theoretical and
              Computational Physics, The University of Hong Kong, Hong Kong, China
         \and
              Department of Astronomy, Nanjing University, Nanjing 210093, China
             }

\abstract{ X-ray transients appeared in optically non-active galactic nuclei have been observed in recent years. The most popular model explaining this kind of phenomena is the conventional tidal disruption model. In this model, when a star moves within the tidal radius of a black hole, part of the star materials will fall into the black hole through an accretion disk, which gives rise to the luminous flare. We propose that the X-ray emission may not necessarily come from radiation of the accretion disk alone. Instead, it may be related to a jet. As the jet travels in the interstellar medium, a shock is produced and synchrotron radiation is expected.  We compared the model light curve and the synchrotron radiation spectrum with the observed data, and find that our model explains the observed light curve and late-time spectrum well. Our model predicts that these transient active galactic nuclei could be sources of the future gamma-ray satellites, e.g. GLAST and the emission r!
 egion will be expanding with time.
}

\date{Received:    }

   \maketitle
%________________________________________________________________

\section{Introduction}

The strong X-ray variations at the galactic nuclei of optically non-active galaxies has been discovered since 1990s \citep[e.g. see][]{badkom96, komgre99, grutho99}. Some of these variations are actually from Seyfert galaxies (e.g. WPVS0007 \citep{grubeu95, grusch07}) while some others are from optically normal, non-active galaxies (e.g. RX J1242.6-1119A) \citep{halgez04}. In this paper we would like to focus on the emissions from normal galaxies. There are common characteristics of these sources. Firstly, all these sources have been bright sources, their X-ray luminosity could go up to about $10^{44}$ erg/s. Secondly, they have shown high level of variability in their X-ray light curve within years. At the `high state', the luminosity of one source could be at least 100 times higher than its luminosity at its `low-state'. Thirdly, most of them have a super soft spectrum during the flare, with effective black body temperatures only about 10 - 100eV  \citep{kombad99, halgez04}. The classic examples which satisfy these characteristics are RX J1624.9+7554, RX J1242.6-1119A, RX J1420+5334, RX J1331-3243 and NGC 5905. Note, however, that NGC 5905 is a starburst galaxy \citep{halgez04}.

Many scenarios have been proposed to explain this phenomena. However, most of them fail to explain some of the observed results. A detailed discussion on these scenarios can be found in \citet{kombad99}. Among all the listed models, the tidal disruption model is the most commonly accepted model and it gives the most satisfactory explanation to the observations by considering the radiation from the disk. However, it still cannot fit the situation perfectly.

In this paper, we try to propose an alternative radiation mechanism which may explain the areas that cannot be explained by the conventional tidal disruption model. We note that when a star is captured by a supermassive black hole at the galactic centre in the tidal disruption model, the star will be accreted to the black hole and a jet may be launched during the accretion process. We propose that the radiation emitted in a shock produced by the mildly relativistic jet can also form an important component in the X-ray flare. In section 2, we discuss the characteristics of some classic flares. In section 3, we will give a more detailed description on the tidal disruption model for completeness. In section 4, we provide a detailed discussion on our jet emission model and some numerical results. Finally, we give a brief conclusion and discussion in section 5.

\section{Typical sources}

This section summarizes the observational results of the commonly accepted candidates of the X-ray flare events satisfying the criterions given in the introduction. We focus on the spectral properties rather than the scale of luminosity variation, which has been discussed in depth already \citep[See, e.g.][]{vauede04, halgez04}. Among all, NGC 5905, RX J1242.6-1119A have been studied in detail while little information is known for RX J1331.9-3243.

\subsection{RX J1242.6-1119A}

This is an interesting source and its detailed observational results can be found in \citet{komgre99} and  \citet{komhal04}. RX J1242.6-1119A was observed two times by ROSAT, with the first time during RASS(ROSAT all-sky survey) during DEC 1990 -- JAN 1991. The second time was a pointed observation with PSPC (position sensitive proportional counter). It was not detected in the RASS (this gives count rate $< 0.015$ cts/s) but it became visible in the pointed observation with the maximum count rate given by 0.3 cts/s \citep{komgre99}. From this detection, it is found that the photon index of this source is $-5.1\pm0.9$ if it is fitted by a power law model($\chi^2_{red} = 1.5$) and gives a temperature of $7 \times 10^5$K when fitted to a blackbody model ($\chi^2_{red} = 0.7$)\citep{komgre99}. This means that the source is very soft.

There were two follow up observations by Chandra (at 9 MAR 2001) and XMM (in 21--22 JUN 2001). The observation by Chandra helped precise localization of the X-ray source.  From the observation by XMM, it is found that the photon index of the source has changed to $-2.5\pm0.2$ \citep{komhal04}. In other words, the source has hardened. Since both Chandra and XMM observations suggest that the detection is a point source, we believe that the detected X-ray was still related to the flare at the nucleus rather than the other parts of the galaxy \citep{komhal04}.

Careful studies in the observation by different telescopes in the optical band suggests that there is no AGN activities in the galactic centre \citep[see, e.g.][]{halgez04}. There is also no radio detection for this source by NRAO VLA Sky Survey(NVSS) \citep{Akom02}.

\subsection{NGC 5905}

NGC 5905 is not non-active since HII lines can be found from the optical spectrum. However, as discussed in \citet{halgez04}, due to the huge variation in luminosity, tidal disruption event is still the favoured model for the flare. Therefore we included this source here for discussion.

NGC 5905 was observed six times by ROSAT, with the first three times observed during RASS, the next two times were pointed observations  by PSPC and the last one by HRI (High Resolution Imager) \citep{badkom96,kombad99}. However, it could only be detected in one of the RASS observations, one of the pointed observations and the HRI observation with count rates found to be 0.6 cts/s, 0.007 cts/s and 0.0007 cts/s respectively  \citep{badkom96,kombad99}. Clearly, there was a huge drop in luminosity. The spectrum has also been found to be hardened. From the observational result of RASS, the photon index is found to be $-4.0\pm0.4$ in the power law model, while in the pointed observation, the photon index become $-2.4\pm0.7$ \citep{badkom96}. 

NGC 5905 is observed again by Chandra in OCT 2002. However, the image is a diffused source, with all photons below 1.5keV \citep{halgez04}. It is impossible for us to determine if the photons positioned at the nuclei of the galaxy came from the flare or from the star burst region. The collected photons are also too few for spectral analysis.

\subsection{RX J1624.9+7554}

RX J1624.9+7554 was searched by ROSAT for two times, but only detected in the RASS in OCT 1990 with a count rate of 0.54 cts/s. The second observation was made in JAN 1992 by PSPC giving the count rate an upper limit of 0.023 cts/s \citep{grutho99}. Several spectral models have been tried to fit the data but only models with a power law component (i.e. power law,  power law free, power law plus a black body spectrum) can give a reasonable fit. The photon index in these models range from $-4.1$ to $-3.3$ \citep{grutho99}. In case the black body model is involved, the temperature is found to be $1.2 \times 10^6$ K \citep{grutho99}.

There was a follow-up observation for this source by Chandra. Since there were only 3 to 4 counts collected in the whole detection period ($\sim 10$ ks) \citep{halgez04}, spectral fitting is not meaningful. However, as argued by \citet{halgez04}, since all photons lies in the range of 0.7 - 4.8 keV, a blackbody spectrum is unlikely and so a power law spectrum can be assumed. 

Similar to RX J1242.6-1119A, studies in the observation by different telescopes in the optical band suggest that there is no AGN activities in the galactic centre \citep[see, e.g.][]{halgez04}. There is also no radio detection for this source by NVSS \citep{Akom02}.

\subsection{RX J1420.4+5334}
The information below is a summary of \citet{gresch00}.
This source was observed four times by ROSAT. However, it was detected only in the second observation, and the other three observations could only give an upper limit to the count rates.
Although there are follow up observations by Chandra, the results have not been published yet.
By comparing the flux of the brightest observation with the other three observations, this source showed a variation of more than 150. The spectral type of this source is unknown since all tested models (i.e. blackbody, disk blackbody and power-law) seems to fit the spectrum equally well. However, it is for sure that it is a soft source since the photon index in the power-law fit is found to be $-5.8$.

Note that optically there are two galaxies within the ROSAT error box, and we cannot single out which one is the host galaxy of the flare. The brighter galaxy, which is likely to be an elliptical or early spiral galaxy, is of redshift $z = 0.147 \pm 0.001$. The fainter one, which seems to be a non-active galaxy, gives the lower limit on the redshift of $z \sim 0.07$. Since the optical observations suggest that its host galaxy is likely to be non-active, it is treated as a candidate for a tidal disruption event. There is no radio detection for this source by NVSS \citep{Akom02}.

\subsection{RX J1331.9-3243}
The information below is a summary of \citet{reigre01}.
This source appeared two times in five ROSAT observations. Its luminosity has increased an order of magnitude within a week. Other details can be found in Table \ref{table1}. It was also non-detectable six month prior and after the flare. The optical spectrum of this source shows no signs of AGN activities so it is treated as a candidate for the tidal disruption event.

\subsection{Summary of the sources}
\begin{table}[htbp]
\caption{A summary of the sources, quoted from \citet{Akom02}, \citet{gresch00}, \citet{reigre01}; \citet{komgre99} } 
\centering 
\begin{tabular}{cccccc} 
\hline
\hline  \\ 
  Galaxy   & z  & k$T_{\rm bb}$  & $L_{\rm x,bb}$ & Photon\\ [2pt]
  && [keV] & [erg/s] &Index\\ [5pt]
\hline
\hline  \\  
  NGC 5905 &  0.011  &  0.06  & 3 $\times$ $10^{42}$ & $-4.0$\\ [3pt] 
  RX J1242.6-1119A &  0.050  &  0.06  & 9 $\times$ $10^{43}$ & $-5.1$\\ [3pt] 
  RX J1624.9+7554 &  0.064  &  0.097  & $\sim$ $10^{44}$ & $-3.3$\\ [3pt] 
  RX J1420.4+5334 &  0.147*  &  0.04  & 8 $\times$ $10^{43}$ & $-5.8$\\ [3pt] 
  RX J1331.9-3243 &  0.051   &    & $\sim 10^{44}$ & $-3.8$\\ [5pt] 
\hline \\ [2pt]
\end{tabular}   
\label{table1}
\\*Please refer to Section 2.4 for discussion on the redshift.
\end{table}  

Table \ref{table1} shows the important properties of these sources. According to \citet{Akom02}, $L_{\rm x,bb}$ given in column 4 are intrinsic luminosities of the flares in $0.1-2.4$ keV band based on the black body model. They should be a lower limit of the peak luminosities since we are most likely not observing the peak of the flares. The photon indexes listed in the column 5 are the one found in the `high state' of the galaxy. 

As a summary, all the five sources listed above exhibit some common characteristics. For example, the sources have shown huge variation in their X-ray luminosity, but they are silent in optical and radio band. They are all point sources and their spectrum can be fitted by the power law model. Moreover, for sources with follow-up observations, their spectrum seems to be hardened after the peaks of the flares.  In general, the flares faded away in a few months to years. The flaring or decaying light curves of the above sources can be found in e.g. \citet{vauede04} and \citet{halgez04}. \citet{halgez04} have shown that the decaying light curve can also be described by the power law model with the power law index given by -5/3. Among all the sources, NGC 5905 has the largest number of observations. Therefore, we will use the data from NGC 5905 for illustration in Section 4.

\section{Review of Tidal Disruption Model}

The ideas of tidal disruption by a supermassive black hole has started as early as 1970s \citep[e.g. see][]{ree88}. It is proposed that the sudden rise in X-ray luminosity of the galactic nuclei of optically non-active galaxies is related to this type of events \citep[e.g. see][]{badkom96, komgre99, grutho99}. We review the classic description of the tidal disruption model very briefly here for completeness. 

Let $M_{BH}$  be the mass of the galactic black hole,  $M_{*}$ and $r_{*}$ be the mass and radius of the disrupted star respectively. According to \citet{ree88}, if a star orbited to a position within the tidal radius $r_t$ of a black hole, where
\begin{equation}
r_t \approx 5 \times 10^{12}\left(\frac{M_{BH}}{10^6M_\odot}\right)^{\frac{1}{3}}\left(\frac{M_{*}}{M_\odot}\right)^{-\frac{1}{3}}\frac{r_*}{r_\odot} \rm cm,
\end{equation}
the star will be heavily distorted. We consider the case with $r_t$ greater than the Schwarzschild radius of the black hole so that only part of the star will be swallowed by the black hole. Conventionally, electromagnetic radiation emitted in three difference processes is considered \citep{kom01}. The processes include (1) interaction among the disrupted, but bounded stellar materials \citep{kimpar99} (2) the accretion process \citep[e.g. see][and the discussion below]{Akom02} (3) the interaction between the unbounded stellar materials with the interstellar medium \citep{khomel96}. %Since process (1) does not give rise to high enough luminosity in the order of months and process (3) is related to a 

According to the numerical simulations done by \citet{ayaliv00}, for a solar type star, about $10\%$ - $50\%$ of the mass will get bounded and eventually accreted into the black hole while the remaining part of the star will be driven away to conserve the angular momentum. As the material accrete to the black hole, radiation is given out due to the hot accretion disk. According to \citet{ulm99}, the black body temperature of accreting a solar-typed star through a thick disk at the tidal radius is given by
\begin{equation}
T_{\rm eff} \approx 2.5 \times 10^5 \left(\frac{M_{BH}}{10^6M_\odot}\right)^{1/12} $ K,$
\label{eq2}
\end{equation}
% isn't this too low for X-ray flare?
 and this will characterize the spectrum of the tidal disruption event. However, more detailed description on the spectrum relies mainly on numerical simulation rather than analytical calculation.
 
\citet{halgez04} have pointed out that the decaying light curve follows the power law of the form 
\begin{equation}
 L_x \propto \left(\frac{t - t_D}{t_0 - t_D}\right)^{-5/3},
 \label{eq3}
\end{equation}
where $t_0$ is the time when the flares started and $t_D$ is the time when the disruption began. \citet{halgez04} have also estimated that for a tidal disruption of a non-rotating star, 
\begin{equation}
(t_0 - t_D) \approx 1.32\times \left(\frac{M_{BH}}{10^6M_\odot}\right)^{1/2}\rm month.
\label{eq4}
\end{equation}
This gives the timescale for the fading of the flare.

\citet{halgez04} have given estimated $t_D$ using equation (\ref{eq3}) and the results are shown in Table \ref{table2}. The third column $t_{obs,p}-t_D$ gives the number of days between the first observation of the flare and $t_D$.

\begin{table}[hpht]
\caption{The parameter $t_D$ of different sources} 
\centering 
\begin{tabular}{ccc} 
\hline
\hline  \\ 
  Galaxy   & $t_D$ [yr]   & $t_{obs,p}-t_D$ [day]\\ [5pt]
\hline
\hline  \\  
  NGC 5905 &   1990.4   &  $\approx50$\\ [3pt] 
  RX J1242.6-1119A &    1991.36   &  $\approx584$\\ [3pt] 
  RX J1624.9+7554 &  1990.7   & $\approx24$ \\ [5pt] 
\hline \\ 
\end{tabular}   
\label{table2}
\end{table}  

In general, the luminosity-time relationship in the observations can be well fitted by equation (\ref{eq3}) \citep[refer to Fig. 4,][]{halgez04}. The estimation of the black hole mass using equation (\ref{eq2}) is also reasonable. \citet{donbra02} has also estimated that the frequency of the tidal disruption event is about $10^{-5}$ per year in every normal galaxy. This frequency also matches with current observation \citep{kom01}. 

%\begin{equation}
%f \approx 10^{-4}M_6^{4/3}\left(\frac{N_*}{10^5pc^{-3}}\right)\left(\frac{\sigma}{100 km s^{-1}}\right)\left(\frac{r_{min}}{r_t}\right)
%\end{equation}

\section{Jet Emission}

\subsection{Motivation}
Although the tidal disruption event explains the observed flares reasonably well, especially that it can provide enough energy for the huge luminosity variations and correct decaying timescale, this pure disk emission model still misses several important points in observations. 

Firstly, a thermal component can usually be found in disk emission. However, according to past analysis of the sources (cf. Section 2) the thermal component can be very insignificant in observed spectrum. Especially, the black body model can only give a poor fit to RX J1624.9+7554 even during the flare. For the sources with follow-up detections by either ROSAT, Chandra or XMM, it can be seen that generally the spectrum hardened and can \textit{only} be described by the power law model. Therefore, it is likely that other radiation mechanisms may involve in the process.

Secondly, according to the fitting results listed in \citet{halgez04}, the peak flux of two flares only occurred about tens of days before their first detections by ROSAT. For a flare which last over months to years, it would be too lucky if we could detect these sources at the starting time. Although one may argue that the flare may have a long peak duration (about 1 year) if the black hole is small enough so that it is easier to catch the peak of the flare \citep{halgez04}, it will then be strange since the luminosities of all the detected sources falls in the follow-up detections and none stay at the peak.

In order to explain the above points, a new radiation mechanism is needed. Moreover, up to now, there is no specific model or simulation presented to explain the observed hardening of the spectrum as time increases. We explore below whether the jet contribution can solve these problems.

\subsection{The Model}

Consider that during an accretion event, a jet can be formed and is ejected from the black hole/ accretion disk. This idea comes from the fact that accreting black holes are seen to be accompanied with jet. One example of such system is the microquasars. Studies on microquasars reveal that these objects behave very differently in their high/low state. 
In their low state, there is evidence of jet emission although the bulk Lorentz factor of the jet is likely to be less than 2  \citep[e.g. see][]{GalFen03}. When they are in the high/soft state, there is evidence that the jet formation is greatly suppressed \citep{fencor99,GalFen03}. However, in their ``very high'' state, the jet reappears again. Unlike the jet seen in the low state, the jet is very powerful and highly relativistic in the ``very high'' state of a microquasar \citep[e.g. see][]{Fen03}. This example shows that transient accreting systems may also be accompanied with powerful jets. 
\citet{wanhu05} have also discussed the jet formation scenario which occurred during stellar capture events by galactic black holes to explain the X-ray cavities in some elliptical galaxies. If jet does exist during a tidal disruption event, the ejected jet will interact with the interstellar medium (ISM) and decelerate accordingly. Synchrotron radiation from jet can explain the power-law decaying light curve and non-thermal spectrum of gamma-ray burst (GRB) afterglow very well \citep[e.g. see][]{pir99, vankou00, grasar02, zhames04, zha07}. Here we try to use the jet emission to explain the non-thermal X-ray flares from normal galactic nuclei.

In the numerical simulation given below, we assume that the jet expands laterally and satisfies four dynamical differential equations \citep{huagou00},

\begin{equation}
\frac{dR}{dt} = \beta c\Gamma(\Gamma + \sqrt{\Gamma^2 -1}),
\end{equation}

\begin{equation}
\frac{dm}{dR} = 2\pi R^2 (1 - cos \theta) n m_p,
\end{equation}

\begin{equation}
\frac{d\Gamma}{dm} = - \frac{\Gamma^2 - 1}{M_{ej} + \epsilon m + 2 (1 - \epsilon) \Gamma m},
\label{dthetadt}
\end{equation}

\begin{equation}
\frac{d\theta}{dt} = \frac{c_s (\Gamma + \sqrt{\Gamma^2 - 1})}{R},
\end{equation}
where $\Gamma$ is the bulk Lorentz factor of the jet, $R$ is the radius of the shock, $\beta=v/c$, where $v$ is the bulk velocity and $c$ is the speed of light in vacuum, $m$ is the swept up mass by the jet, $\theta$ is the half opening angle of the jet, $n$ is the environment density, $M_{ej}$ is the ejected mass, $\epsilon$ is the radiative efficiecy and $c_s$ is the comoving sound speed. 

Since the jet is probably mildly relativistic, as it moves into the ISM, a shock front is formed and synchrotron radiation will be resulted. According to \citet{ryblig79}, the synchrotron radiation power at frequency $\nu^{\prime}$ in the jet comoving frame is given by 
\begin{equation}
P'(\nu') = \frac{\sqrt{3}e^3B'}{m_ec^2} \int_{\gamma_{\rm e,min}}^{\gamma_{\rm e,max}}\left(\frac{dN'_e}{d\gamma_e}\right)F\left(\frac{\nu'}{\nu'_c}\right) d\gamma_e,
\end{equation}
where e is the charge of the electron, $B'$ is the comoving magnetic field, $m_e$ is the electron mass, $\gamma_e$ is the electron Lorentz factor. The critical frequency $\nu'_c$, the minimum and maximum Lorentz factor $\gamma_{\rm e,min}$, $\gamma_{\rm e,max}$ and the function $F(x)$ is given by the equations  \citep{huagou00}
\begin{equation}
\nu'_c=\frac{3\gamma_e^2 eB'}{4\pi m_e c},
\end{equation}

\begin{equation}
\gamma_{\rm e,min} =\epsilon_e(\Gamma - 1)\frac{m_p(p-2)}{m_e(p-1)} +1 \mbox{    for $p > 2$,}
\end{equation}

\begin{equation}
\gamma_{\rm e,max} = 10^8(B' /1G)^{-1/2},
\end{equation}

\begin{equation}
F(x) = x \int_x^{+ \infty} K_{5/3}(k)dk,
\end{equation}
where  $\epsilon_e$ is the fraction of energy carried by the shocked electrons to proton energy, $p$ is electron energy distribution index (see discussion below) which is typically about  $2-3$ and $K_{5/3}(k)$ is the Bessel function.
Finally, $\frac{dN'_e}{d\gamma_e}$ is the electron energy distribution. Its expression is determined by the relationship between $\gamma_{\rm e,min}$, $\gamma_{\rm e,max}$ and $\gamma_{c}$, where $\gamma_{c}$ is defined by \citep{sarpir98}
\begin{equation}
\gamma_c = \frac{6\pi m_e c}{\sigma_T\Gamma B'{}^2 t}.
\end{equation}
According to \citet{huache03}, the expression of $\frac{dN'_e}{d\gamma_e}$  is given by

Case 1:  $\gamma_c \leq \gamma_{\rm e,min}$, 
\begin{equation}
\frac{d N' _{e}}{d \gamma_e} = C_1 (\gamma_e -1)^{-(p+1)},          \mbox{   where}
\end{equation}

\begin{equation}
C_1 = \frac{p}{(\gamma_{\rm e,min}-1)^{-p} - (\gamma_{\rm e,max}-1)^{-p}}N_e;
\end{equation}

Case 2: $\gamma_{\rm e,min} < \gamma_c \leq \gamma_{\rm e,max}$,

\begin{equation}
\frac{d N' _{e}}{d \gamma_e} = \left\{ \begin{array}{ll}
C_2(\gamma_e - 1)^{-p} & \mbox{for $\gamma_{\rm e,min} \leq \gamma_e \leq \gamma_c$},\\
C_3(\gamma_e - 1)^{-(p+1)} & \mbox{for $\gamma_c < \gamma_e \leq \gamma_{\rm e,max}$},
\end{array} \right.
\end{equation}

where
\begin{eqnarray}
\lefteqn{}
  C_3        &  = & \left[ \frac{(\gamma_{\rm e,min} - 1)^{1-p} - (\gamma_c - 1)^{1-p}}{(p-1)(\gamma_c - 1)} \right. \nonumber\\
                  &     & \left. \mbox{} + \frac{(\gamma_c - 1)^{-p} - (\gamma_{\rm e,max} - 1)^{-p}}{p}  \right]^{-1}N_e,
\end{eqnarray}

\begin{equation}
C_2  = \frac{C_3}{\gamma_c - 1};
\end{equation}

Case 3:  $\gamma_c > \gamma_{\rm e,max}$,

\begin{equation}
\frac{d N' _{e}}{d \gamma_e} = C_4 (\gamma_e - 1)^{-p},          \mbox{   where}
\end{equation}

\begin{equation}
C_4 = \frac{p - 1}{(\gamma_{\rm e,min} - 1)^{1-p} - (\gamma_{\rm e,max} - 1)^{1-p}} N_e.
\end{equation}

In order to calculate the flux at frequency $\nu$ in the observers frame, we first assume that the radiation is isotropic in the comoving frame, i.e.
\begin{equation}
\frac{dP'(\nu')}{d\Omega'} = \frac{P'(\nu')}{4\pi}.
\end{equation}
Let $\Theta$ be the angle between the velocity of the emitting material and the line of sight and define $\mu = \cos\Theta$. Then, by \citet{ryblig79},
\begin{equation}
\frac{dP(\nu)}{d\Omega} = \frac{1}{\Gamma^3(1 - \beta \mu)^3}\frac{P'(\nu')}{4\pi} ,
\end{equation}
and the relationship between $\nu$ and $\nu'$ is given by
\begin{equation}
\nu = \frac{\nu'}{\Gamma (1 - \beta \mu)}.
\end{equation}
Therefore, the observed flux at frequency $\nu$ from a source with luminosity distance $D_L$  is given by \citep{huadai00}
\begin{equation}
S_\nu = \frac{1}{4\pi D_L^2}\frac{P'(\Gamma (1 - \beta \mu)\nu)}{\Gamma^3(1 - \beta \mu)^3}. 
\end{equation}

Although this mechanism is similar to the one responsible for the afterglow of a GRB, the initial conditions of the jet emitted during the accretion to the supermassive black hole at the galactic centre can be very different from that of a typical GRB.  \citet{cheche06} and \citet{luche06} have estimated the maximum energy $\Delta{E_p}$ carried away by the relativistic protons in the jet from the accretion disk as
\begin{equation}
\Delta{E_p} \sim 6 \times 10^{52}(\eta_p/10^{-1}) (M_*/M_{\odot}) \mbox{erg},
\end{equation}
where $\eta_p$ is the conversion efficiency  from accretion power into the energy of jet motion and $M_*$ is the mass of the disrupted star.
>From this equation, we assume that the ejected energy is of the order of about $10^{52}$ ergs. This estimation of energy in the jet is consistent with the observed high energy data in the galactic centre. In case that $\eta_p$ is much smaller than 0.1, we may still be able to observe the jet emission if a more massive star is captured. For example, \citet{luche06} have studied the situation when a red giant star is captured by the black hole.

The initial Lorentz factor $\Gamma$ of the jets ejected from the galactic black holes should also be much lower than the jets responsible for the GRB afterglows. The jet's initial Lorentz factor is $>100$ for a typical GRB. However, for a typical AGN, the jet's initial Lorentz factor is only of the order of several. For example, from \citet{chefan99}, the Doppler factors of the studied AGNs are all found to be less than 10. For small viewing angle, the Doppler factors and the Lorentz factors of the jets should be of the same order. \citet{yuamar02} has also estimated the early bulk Lorentz factor of NGC 4258 (a low luminosity AGN) is about 3. Since during the flare emission, the situation in the galactic centre may be similar to an AGN, we should restrict the jet initial Lorentz factor to be of the order of several.

\subsection{The Results of Numerical Simulation}

We give the numerical simulation of the jet model here for illustrating the compatibility between the jet model and the flare events. We have used the jet program initially for simulating the afterglow light curve of GRB and the details of the program can be found in \citet{huadai00, huagou00} and \citet{huache03}. However, to suit the environment in the galactic centre, we have used the input parameters as follow: The most important input parameters are the jet energy $E_{0,jet} = 10^{52}$ ergs, the initial bulk Lorentz factor $\Gamma = 2.5$ and the initial jet opening angle $\theta_0 = 0.3$. Other parameters include, the electron energy distribution index $p = 2.2$, the fraction $\epsilon_B = 0.01$ of magnetic energy density to thermal energy density, the fraction $\epsilon_e=0.1$ of energy carried by the shocked electrons to proton energy and the environmental density $n = 1000$ cm$^{-3}$. These parameters are used for all the simulation results below unless otherwise sta!
 ted. Note that they in principle can be time and radius dependent, we assume that they are constant for simplicity in illustration. We also assume that we are looking at the jet along the central axis and the source is 50 Mpc away. In this simplified model, we also assumed that all the ejecta are ejected at once and there is only one ejection, while in many GRB afterglow model, there can be continuous energy injections or several discrete energy injections. We calculated the flux in the energy range of 0.1 -- 2.4 keV for comparison of the results in literatures.

\begin{figure}[htbp]
\begin{center}
\includegraphics[width = 3.5in]{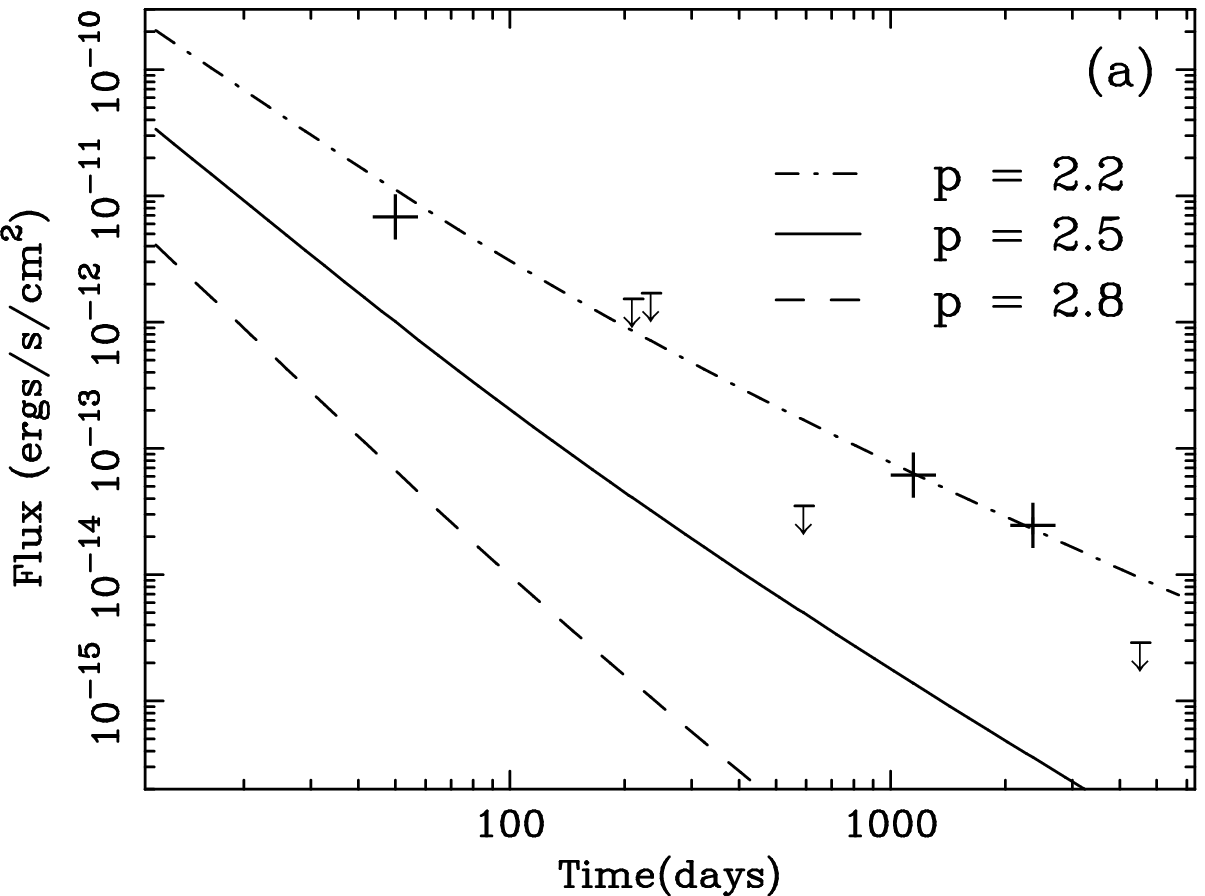}
\includegraphics[width = 3.5in]{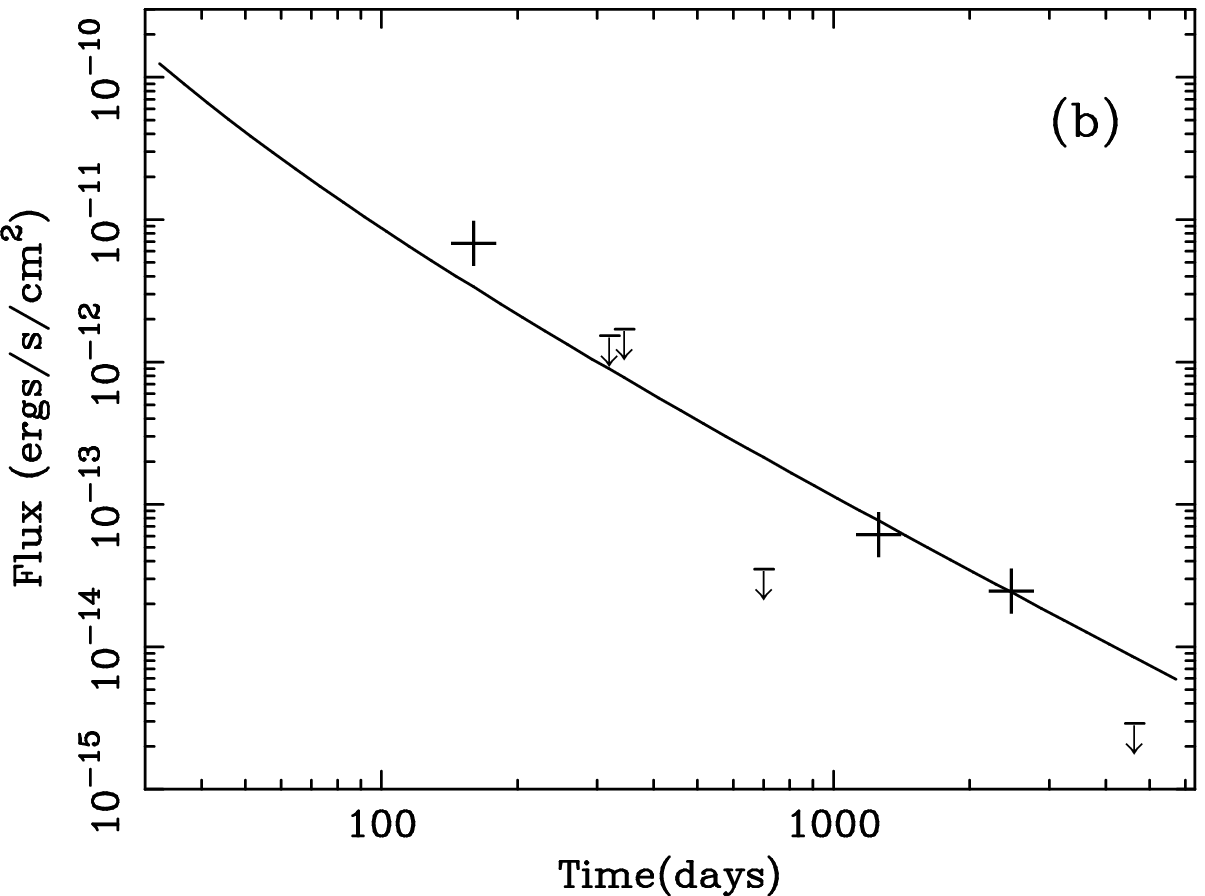}
\caption{ (a) Observed (0.1 - 2.4 keV) X-ray flux in NGC 5905 together with the simulation results. The data points for NGC 5905 is extracted from \citet{badkom96,halgez04} and the trigger time is assumed to be $\sim50$ days before the first observation; In our theoretical calculation, we have taken $E_{\rm 0,jet} = 3 \times 10^{51}$ ergs, $\Gamma$=5, $\epsilon_B = 0.001$ and $\epsilon_e=0.06$. (b) Same as (a), but the trigger time is assumed to be $\sim160$ days before the first observation. with the simulated light curve is given by another set of model parameters ($E_{\rm 0,jet} = 10^{52}$ ergs, $\Gamma=8.3$, $\epsilon_B = 0.08$, $\epsilon_e=0.5$ and $p=2.4$). This illustrates that different parameter sets allow for different trigger time.}
\label{Fig1}
\end{center}
\end{figure}

Our theoretical light curves are shown in Figure \ref{Fig1}. From our simulation results, when p = 2.2, we found that the light curve follows the relationship $L_x \propto t^{-\alpha}$, where $\alpha$ changes from $\sim2$ at the beginning of the flare to about $\sim 1.3$ at the later stage (i.e. about a few years after the burst). We can change the power law index $\alpha$ to a broad range easily by varying the electron distribution index $p$. Observed data from NGC 5905 are also included for comparison. Note that the intrinsic trigger time of this event is unknown. In Figure \ref{Fig1}a, we assume that the trigger time is about 50 days before the first observation, which is the trigger time suggested by \citet{halgez04}. We see that the dash-dotted line can fit the observations well. In Figure \ref{Fig1}b, the trigger time is assumed to be about 160 days before the first observation. Again we see that the observations can be well fitted by changing the model parameters acco!
 rdingly. As a conclusion, we found that even the jet model alone can explain the observed light curve of NGC 5905 reasonably well in both case.

Furthermore, the above results also illustrate that the jet model relaxes the tight constraint on when the flare begin since Figure \ref{Fig1}a and \ref{Fig1}b together show that different set of model parameters allows the starting time of the flare to shift by more than a hundred days. The flexibility of the jet model comes from the fact that the realistic light curves of the flares actually depend on many parameters, such as $p$ and $\epsilon_e$, these parameters can pick a range of values and it is possible that these values are function of time $t$ and position $R$.  Note that if the actual trigger time is changed, it mainly changes the time position of the first data point in Figure \ref{Fig1}, but it does not affect the positions of the other data points significantly. 

\begin{figure}[htbp]
\begin{center}
\includegraphics[width = 3.5in]{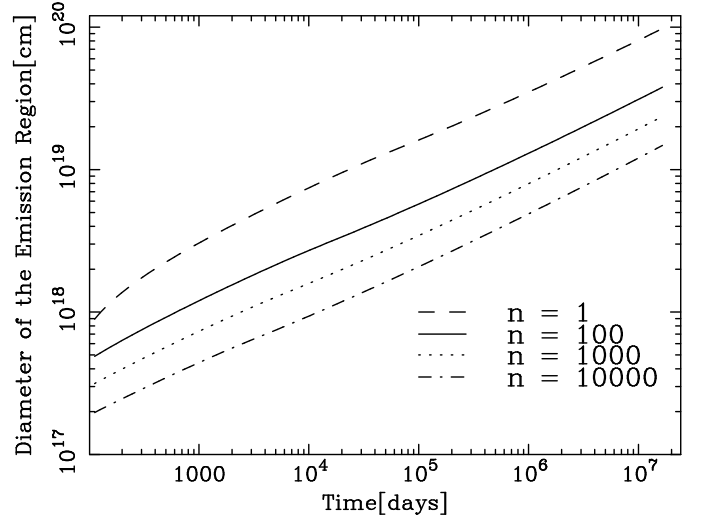}
\caption{ Time evolution of emission size in different ISM density [in cm$^{-3}$]}
\label{Fig2}
\end{center}
\end{figure}

In our model, the emission size is much larger than an accretion disk. Time evolution of the size of the emission regions is shown in Figure \ref{Fig2}. However, by the angular resolution of the current X-ray telescope, emission size is still non-resolvable. Therefore the current observational results cannot be used to distinguish if our model is correct as long as the source is point-like. 

\begin{figure}[htbp]
\begin{center}
\includegraphics[width = 3.5in]{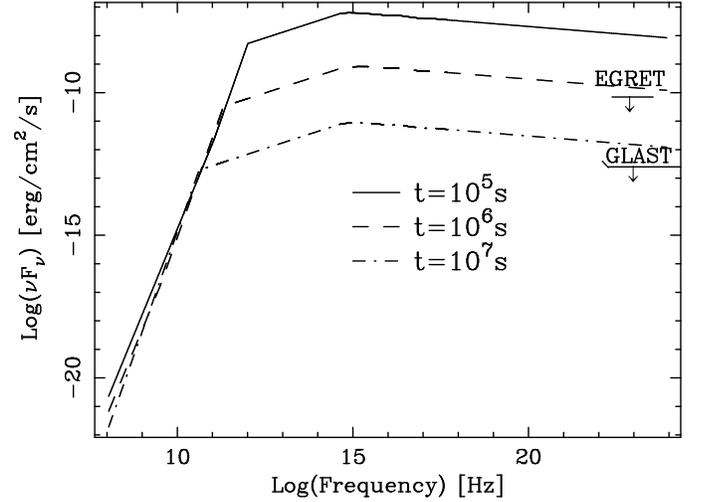}
\caption{ Broad band spectrum predicted by the jet model. The used parameters are $E_{0,jet} = 10^{52}$ ergs, $\Gamma = 5$, $\theta_0 = 0.3$, $p = 2.2$, $\epsilon_B = 0.0001$, $\epsilon_e=0.1$, $n = 1000$ cm$^{-3}$.}
\label{Fig4}
\end{center}
\end{figure}

Figure \ref{Fig4} shows a typical broad band spectrum of synchrotron radiation as described in, for example, \citet{grasar02}. The synchrotron spectrum is a broken power law spectrum with the photon index in the X-ray band given by $-(p/2+1)$, where p is the energy distribution index of electron inside the jet. If we take $p=2.2$, the photon index should be about $-2.1$, which matches spectrum index at the late stage. This gives evidence that at least part of the radiation from the flare is from the jet rather than solely from the accretion disk.

\section{Discussion and Conclusion}

As a conclusion, in this paper we have summarized the observed features of some tidal eruption candidates and concluded that the pure accretion disk model cannot explain the radiation properties from these sources completely, especially at the late stage of emission. In order to explain the situation more throughly, we have proposed the jet model.  We suggest that when a star is captured by a supermassive black hole at the galactic centre, the star will be accreted to the black hole and a jet will be emitted during the accretion process. The jet will then decelerate in the ISM, and synchrotron radiation is given out in the shock emission region. This form part of the radiation in the X-ray transient event. This jet model is able to explain the late non-thermal spectrum and also relaxes the tight constraints on the trigger times of the events.

In our calculations, we take the model parameters by referring to the situation in AGNs, as discussed in Section 4.2, since the parameter values are unknown for these flaring galaxies. In other words, we assume that during the flaring states, these ``normal galaxies'' are similar to the AGNs. This is the best choice that we can make currently. In the future, when more is known for these flaring galaxies, the model parameters may be refined accordingly. Our simulation results show that the jet model alone can explain the observed light curve of NGC 5905 reasonably well. The jet model also predicts larger emission region, though current X-ray telescopes do not have good enough angular resolution to test our prediction. 

\begin{figure}[htbp]
\begin{center}
\includegraphics[width = 3.5in]{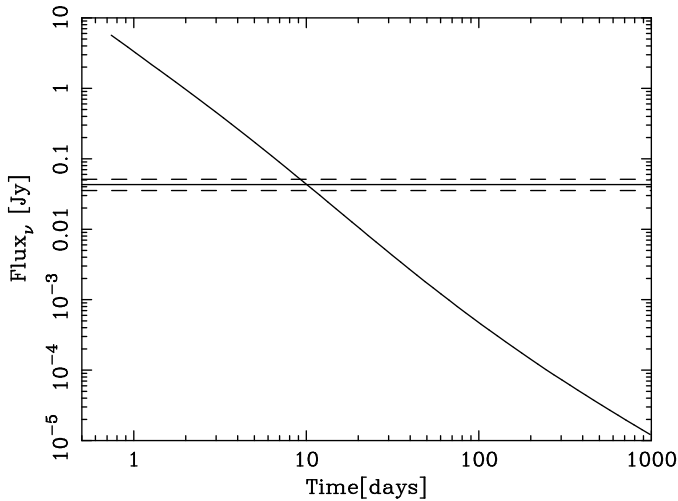}
\caption{Optical light curve in B band of NGC5905 predicted by the jet model, using the same parameters used in Figure 1a. The Solid horizontal line represents the observed optical background (Data taken from Simbad, catalog VII/1B) and the dashed horizontal lines represent the allowed variations based on the results given in \citet{kombad99}. }
\label{Fig5}
\end{center}
\end{figure}

Although the model predicts emission in all energy bands, we may not be able to detect radiation in the optical band. The main reason for the non-detection may be the strong optical background radiation. Note that in the jet model, the jet is emitted at the galactic centre, which is usually optically bright. Again, NGC 5905 is used as the example for illustration (see Figure \ref{Fig5}). This is the only source with optical observation close to the detected peak of the flare. \cite{kombad99} concluded that during the flare, the optical magnitude of NGC 5905 does not varies by more than 0.2 magnitude. This is illustrated by the two dashed lines in Figure \ref{Fig5}, which has plotted the predicted flux in B band using the same parameters used in Figure \ref{Fig1}a. For this particular set of model parameters, the model prediction is consistent with Fig 2 in \citet{kombad99} as long as the trigger time at least 10 days between two consecutive optical observations. This example!
  illustrates that for galaxies with optical observations close to the start of the flare, the optical observations may help in constraining the trigger time. For the other flaring sources, the situation was similar except that the optical observations were done years after the peak of the flares, therefore it is even more difficult for us to observe the increase in optical flux.

Another reason may be related to the dust extinction. There is evidence the dust extinction is very serious in galactic nuclei, as in our Milky Way \citep[e.g.][]{scosto03}. In general, dust extinction is important in both the optical and x-ray band, however, the way which the dust affects the optical and x-ray band is not the same. According to \citet{weidra01}, there are evidence that different galaxies may have different grain sizes and it is found that galaxies may have large grain sizes if the optical extinction is strong. In the X-ray band, the grain size is not so important to the degree of dust extinction. Therefore, it may be the case that the flaring galaxies are of large grain size if the non-detection of the optical band is due to dust extinction. Moreover, if the flaring galaxy has a gas/dust ratio that is heavily sub-Galactic, then the optical emission will also be seriously obscured while X-rays are less affected, since X-ray photons are mainly absorbed by gas.

In the radio band, we find that emission is very weak in general (c.f. Fig. \ref{Fig4}). We will not have any detection in the radio band unless the source is very near us. However, future radio telescope (e.g. Atacama Large Millimeter Array (ALMA)) may be able to detect the size of the emission region, by which we can differentiate the region from the disk or from the shock emission region.
The sources are also non-detected in the gamma-ray band, but this is simply because the predicted gamma-ray radiation is too weak for current detection unless they are observed at the very early part of the flare. The thresholds of EGRET and GLAST included in Figure \ref{Fig4} were taken from \citet{chezha04} and \citet{mcemos04}. With the high sensitivity of GLAST, which will be launched at the end of 2007, we should be able to detect the increase in gamma-ray flux.

We noted also that, for most of the identified flares, the spectrum during the flare is extremely soft, which is not a consequence of synchrotron radiation. In these cases, we believe that the synchrotron radiation in the jet is still involved in the radiation, but may not be the dominated mechanism initially. However, as times goes on, the spectrum of the flares become hardened and the photon indexes matched with the typical values given in synchrotron radiation. By this time, synchrotron radiation may become dominant.

Our jet model allows rebrightening or fast variability to occur in the X-ray light curve. For example, when there is a density jump in the ISM or when there is a second injection of energy or materials into the jet, there could be a sudden relatively small scale rise in luminosity in the light curve \citep[e.g.][]{huache06}. However, it is unlucky that the current observational data are only sparsely sampled so that these fast variabilities have not been revealed. 
Moreover, there could be many different ways that a jet can evolve in the ISM. As an example, the geometrical shape of the jet may be cylindrical, but not conical as assumed in our current study. Actually, cylindrical jets have been observed in a few radio galaxies. The emission from a conical jet and a cylindrical jet can be very different \citep{chehua01}. 
It worths further investigation on which kind of jet works better in explaining the X-ray flare events. The results can help us understand the environment of the galactic centres.

\begin{acknowledgements}
We thank V. Dogiel, S. Komossa, P. Predehl and A. Li for useful discussion. We also thank the referee for the detailed comments. A. Y. L. Wong \& K. S. Cheng are supported by a RGC grant of the Hong Kong Government. Y. F. Huang was supported by National Natural Science Foundation of China (grants 10625313 and 10221001).
\end{acknowledgements} 

%\bibliography{Astrophy}

\begin{thebibliography}{43}
\expandafter\ifx\csname natexlab\endcsname\relax\def\natexlab#1{#1}\fi

\bibitem[{Ayal {et~al.}(2000)Ayal, Livio, \& Piran}]{ayaliv00}
Ayal, S., Livio, M., \& Piran, T. 2000, \apj, 545, 772

\bibitem[{Bade {et~al.}(1996)Bade, Komossa, \& Dahlem}]{badkom96}
Bade, N., Komossa, S., \& Dahlem, M. 1996, \aap, 309, L35

\bibitem[{Cheng {et~al.}(2006)Cheng, Chernyshov, \& Dogiel}]{cheche06}
Cheng, K.~S., Chernyshov, D.~O., \& Dogiel, V.~A. 2006, \apj, 645, 1138

\bibitem[{Cheng {et~al.}(1999)Cheng, Fan, \& Zhang}]{chefan99}
Cheng, K.~S., Fan, J.~H., \& Zhang, L. 1999, \aap, 352, 32

\bibitem[{Cheng {et~al.}(2001)Cheng, Huang, \& Lu}]{chehua01}
Cheng, K.~S., Huang, Y.~F., \& Lu, T. 2001, \mnras, 325, 599

\bibitem[{Cheng {et~al.}(2004)Cheng, Zhang, Leung, \& Jiang}]{chezha04}
Cheng, K.~S., Zhang, L., Leung, P., \& Jiang, Z.~J. 2004, \apj, 608, 418

\bibitem[{Donley {et~al.}(2002)Donley, Brandt, Eracleous, \& Boller}]{donbra02}
Donley, J.~L., Brandt, W.~N., Eracleous, M., \& Boller, T. 2002, \apj, 124,
  1308

\bibitem[{Fender(2003)}]{Fen03}
Fender, R. 2003, astro-ph/0303339

\bibitem[{Fender {et~al.}(1999)Fender, Corbel, Tzioumis, McIntyre,
  Campbell-Wilson, Nowak, Sood, Hunstead, Harmon, Durouchoux, \&
  Heindl}]{fencor99}
Fender, R., Corbel, S., Tzioumis, T., {et~al.} 1999, \apj, 519, L165

\bibitem[{Gallo {et~al.}(2003)Gallo, Fender, \& Pooley}]{GalFen03}
Gallo, E., Fender, R.~P., \& Pooley, G.~G. 2003, \mnras, 344, 60

\bibitem[{Granot \& Sari(2002)}]{grasar02}
Granot, J. \& Sari, R. 2002, \apj, 568, 820

\bibitem[{Greiner {et~al.}(2000)Greiner, Schwarz, Zharikov, \& Orio}]{gresch00}
Greiner, J., Schwarz, R., Zharikov, S., \& Orio, M. 2000, \aap, 362, L25

\bibitem[{Grupe {et~al.}(1995)Grupe, Beuerman, Mannheim, Thomas, Fink, \&
  de~Martino}]{grubeu95}
Grupe, D., Beuerman, K., Mannheim, K., {et~al.} 1995, \aap, 300, L21

\bibitem[{Grupe {et~al.}(2007)Grupe, Schady, Leighly, Komossa, O'Brien, \&
  Nousek}]{grusch07}
Grupe, D., Schady, P., Leighly, K.~M., {et~al.} 2007, 133, 1988

\bibitem[{Grupe {et~al.}(1999)Grupe, Thomas, \& Leighly}]{grutho99}
Grupe, D., Thomas, H.-C., \& Leighly, K.~M. 1999, \aap, 350, L31

\bibitem[{Halpern {et~al.}(2004)Halpern, Gezari, \& Komossa}]{halgez04}
Halpern, J.~P., Gezari, S., \& Komossa, S. 2004, \apj, 604, 572

\bibitem[{Huang \& Cheng(2003)}]{huache03}
Huang, Y.~F. \& Cheng, K.~S. 2003, \mnras, 341, 263

\bibitem[{Huang {et~al.}(2006)Huang, Cheng, \& Gao}]{huache06}
Huang, Y.~F., Cheng, K.~S., \& Gao, T.~T. 2006, \apj, 637, 873

\bibitem[{Huang {et~al.}(2000{\natexlab{a}})Huang, Dai, \& Lu}]{huadai00}
Huang, Y.~F., Dai, Z.~G., \& Lu, T. 2000{\natexlab{a}}, \mnras, 316, 943

\bibitem[{Huang {et~al.}(2000{\natexlab{b}})Huang, Gou, Dai, \& Lu}]{huagou00}
Huang, Y.~F., Gou, L.~J., Dai, Z.~G., \& Lu, T. 2000{\natexlab{b}}, \apj, 543,
  90

\bibitem[{Khokhlov \& Melia(1996)}]{khomel96}
Khokhlov, A. \& Melia, F. 1996, \apj, 457, L61

\bibitem[{Kim {et~al.}(1999)Kim, Park, \& Lee}]{kimpar99}
Kim, S.~S., Park, M., \& Lee, H.~M. 1999, \apj, 519, 647

\bibitem[{Komossa(2001)}]{kom01}
Komossa, S. 2001, astro-ph/0109441

\bibitem[{Komossa(2002)}]{Akom02}
---. 2002, astro-ph/0209007

\bibitem[{Komossa \& Bade(1999)}]{kombad99}
Komossa, S. \& Bade, N. 1999, \aap, 343, 775

\bibitem[{Komossa \& Greiner(1999)}]{komgre99}
Komossa, S. \& Greiner, J. 1999, \aap, 349, L45

\bibitem[{Komossa {et~al.}(2004)Komossa, Halpern, Schartel, Hasinger,
  Santos-Lleo, \& Predehl}]{komhal04}
Komossa, S., Halpern, J., Schartel, N., {et~al.} 2004, \apj, 603, L17

\bibitem[{Lu {et~al.}(2006)Lu, Cheng, \& Huang}]{luche06}
Lu, Y., Cheng, K.~S., \& Huang, Y.~F. 2006, \apj, 641, 288

\bibitem[{McEnery {et~al.}(2004)McEnery, Moskalenko, \& Ormes}]{mcemos04}
McEnery, J.~E., Moskalenko, I.~V., \& Ormes, J.~F. 2004, in Cosmic Gamma-Ray
  Sources, ed. K.~S. Cheng \& G.~E. Romero (Dordrecht: Kluwer Academic
  Publishers), 361 -- 395

\bibitem[{Piran(1999)}]{pir99}
Piran, T. 1999, \physrep, 314, 575

\bibitem[{Rees(1988)}]{ree88}
Rees, M.~J. 1988, \nat, 333, 523

\bibitem[{Reiprich \& Greiner(2001)}]{reigre01}
Reiprich, T.~H. \& Greiner, J. 2001, in Black Holes in Binaries and Galactic
  Nuclei: Diagnostics, Demography and Formation, ed. L.~Kaper, E.~van~den
  Heuvel, \& P.~Woudt, ESO Workshop (Springer-Verlag), 168 -- 169

\bibitem[{Rybicki \& Lightman(1979)}]{ryblig79}
Rybicki, G.~B. \& Lightman, A.~P. 1979, Radiative processes in astrophysics
  (New York: Wiley-Interscience)

\bibitem[{Sari {et~al.}(1998)Sari, Piran, \& Narayan}]{sarpir98}
Sari, R., Piran, T., \& Narayan, R. 1998, \apj, 497, L17

\bibitem[{Scoville {et~al.}(2003)Scoville, Stolovy, Rieke, Christopher, \&
  Yusef-Zadeh}]{scosto03}
Scoville, N.~Z., Stolovy, S.~R., Rieke, M., Christopher, M., \& Yusef-Zadeh, F.
  2003, \apj, 594, 294

\bibitem[{Ulmer(1999)}]{ulm99}
Ulmer, A. 1999, \apj, 514, 180

\bibitem[{van Paradijs {et~al.}(2000)van Paradijs, Kouveliotou, \&
  Wijers}]{vankou00}
van Paradijs, J., Kouveliotou, C., \& Wijers, R. A. M.~J. 2000, ARA\&A, 38, 379

\bibitem[{Vaughan {et~al.}(2004)Vaughan, Edelson, \& Warwick}]{vauede04}
Vaughan, S., Edelson, R., \& Warwick, R.~S. 2004, \mnras, 349, L1

\bibitem[{Wang \& Hu(2005)}]{wanhu05}
Wang, J.~M. \& Hu, C. 2005, \apj, 630, L125

\bibitem[{Weingartner \& Draine(2001)}]{weidra01}
Weingartner, J.~C. \& Draine, B.~T. 2001, \apj, 548, 296

\bibitem[{Yuan {et~al.}(2002)Yuan, Markoff, Falcke, \& Biermann}]{yuamar02}
Yuan, F., Markoff, S., Falcke, H., \& Biermann, P.~L. 2002, \aap, 391, 139

\bibitem[{Zhang(2007)}]{zha07}
Zhang, B. 2007, ChJAA, 7, 1

\bibitem[{Zhang \& M\'esz\'aros(2004)}]{zhames04}
Zhang, B. \& M\'esz\'aros, P. 2004, Int. J. Mod. Phys. A, 19, 2385

\end{thebibliography}

\end{document}